# Low costs na aviação: importância e desdobramentos

Bruno Felipe de Oliveira
Alessandro V. M. Oliveira✈
Instituto Tecnológico de Aeronáutica, São José dos Campos, Brasil
✈ Autor correspondente. Instituto Tecnológico de Aeronáutica. Praça Marechal Eduardo Gomes, 50. 12.280-250 - São José dos Campos, SP - Brasil.
E-mail: alessandro@ita.br.

*Resumo*: Este trabalho tem como objetivo discutir os impactos de uma companhia aérea de baixo custo no mercado do transporte aéreo, e especialmente apresentar as descobertas mais recentes da literatura especializada na área. Para isso foram selecionados e analisados diversos trabalhos sobre esse tema que foram publicados desde o ano de 2015. A partir dessa análise foi possível categorizar os principais tópicos discutidos nos trabalhos em cinco grupos, que são: (i) os impactos de uma companhia aérea de baixo custo nas companhias aéreas concorrentes; (ii) os impactos nos aeroportos; (iii) impactos gerais na demanda por transporte aéreo; (iv) efeitos no processo de escolha dos passageiros; e (v) efeitos gerais em uma região geográfica.

*Palavras-chave*: transporte aéreo, companhias aéreas, aeroportos.

## I. Introdução

Desde a desregulamentação da aviação civil nos EUA em 1978, o crescimento de companhias aéreas de baixo custo (do inglês low cost carriers, ou LCCs) vem atraindo atenção de diversos participantes desse setor. Isso pode ser explicado pelo impacto que esses tipos de companhias aéreas têm causado na expansão da indústria da aviação. Em 2015, as transportadoras de baixo custo transportaram mais de 980 milhões de passageiros, o que representava à época 28% do total de passageiros em voos programados no mundo .

Esse número é atingido pelo modelo de negócio "baixo custo e baixa tarifa", no qual as companhias aéreas oferecem serviços limitados a seus passageiros por meio da estratégia de desagregação de serviços (unbundling), em que o serviço essencial de transporte aéreo é separado de outros serviços considerados adicionais, como despacho de bagagens, refeição a bordo, sistema de entretenimento, seleção antecipada de assentos, entre outros. Ao adotar a estratégia de unbundling e/ou aumentando a sua eficiência operacional por meio da alta utilização de aeronaves e frota padronizada, a companhia aérea consegue reduzir o custo de suas operações, o que pode resultar na redução do preço de suas passagens. Ao aplicar esse modelo de negócios, a Southwest Airlines, uma das maiores companhias aéreas de baixo custo do mundo e uma referência mundial nesse segmento, é responsável unicamente por bilhões de economia ao ano para o consumidor em passagens de voos domésticos, o que na prática se traduz na redução do preço médio das passagens e o subsequente aumento na procura por viagens aéreas (Beckenstein e Campbell, 2017). Com essa estratégia, as companhias aéreas são capazes de atender àqueles passageiros sensíveis ao preço, assim expandindo o setor de viagens aéreas .

Desde a Lei de Desregulamentação da Aviação Civil nos EUA, os estudos acadêmicos têm se preocupado com a concorrência no mercado de companhias aéreas com perguntas sobre preços e entrada de companhias aéreas em rotas. Vários estudos foram realizados nessas áreas desde a desregulamentação dos EUA, e o número de pesquisas aumentou ainda mais quando outras regiões do mundo iniciaram sua própria desregulamentação do mercado da aviação civil. Uma das linhas de pesquisa está preocupada com os efeitos da entrada e as respostas às entradas de LCCs por seus concorrentes; nesse tópico, a literatura descobriu que uma companhia aérea de baixo custo efetivamente reduz as tarifas médias quando ela entra no mercado ou mesmo quando ela apenas ameaça entrar em uma rota (Windle e Dresner, 1999; Morrison, 2001; Goolsbee e Syverson, 2008; Brueckner et al. 2013).

Nesse contexto, o presente trabalho tem como objetivo primário lançar novas luzes e avaliar o que está sendo discutido recentemente pelos pesquisadores sobre os efeitos das LCCs não apenas nas tarifas aéreas, mas em diferentes participantes da indústria da aviação civil. Para isso, este trabalho revisará as publicações recentes sobre o assunto, discutirá suas principais conclusões e apontará o que mais poderá ser explorado sobre esse tema no futuro.

## II. Companhias aéreas de baixo custo

A expansão do setor aéreo está intimamente ligada ao crescimento do modelo de negócios de baixo custo. De acordo com a indústria, uma companhia aérea de baixo custo em geral possui as seguintes características: serve rotas de curta distância; usa aeroportos regionais ou secundários; opera rotas ponto a ponto; possuem programas limitados de fidelização de clientes (ou não possuem); serviços limitados aos passageiros; alta proporção de vendas de passagens feitas pela Internet; alta utilização da frota de aeronaves; e frota padronizada. Quando uma companhia aérea opera seguindo algumas (ou todas) características citadas ela diminui seus custos operacionais, permitindo então a adoção de estratégia de baixa tarifa. Reduzindo suas tarifas, uma LCC pode atender passageiros sensíveis ao preço, que de outra forma utilizariam outros modais de transporte ou até deixariam de viajar. Esse aumento na demanda por voos como consequência de uma queda nas tarifas aéreas é conhecido na indústria como "Efeito Southwest".

Este termo foi cunhado por Bennet e Craun (1993) do Departamento de Transporte dos EUA para descrever o aumento nas viagens aéreas que resultou da entrada da Southwest Airlines em novos mercados após a desregulamentação do mercado americano de aviação civil. Segundo eles, o aumento na demanda de viagens aéreas ocorre em três etapas. Primeiro, a entrada da Southwest aumenta a oferta de viagens aéreas em um mercado e, ao mesmo tempo, oferece tarifas mais baixas que seus concorrentes; segundo, as empresas aéreas concorrentes reduziriam suas próprias tarifas para permanecerem competitivas e evitarem perder passageiros para a Southwest; e terceiro, devido à pressão para baixo nos preços, causada pelo aumento da oferta de voos e pela concorrência com uma low cost, a procura por viagens aéreas aumenta naquele mercado com a presença da Southwest.

Desde então, vários estudos investigaram o Efeito Southwest ao longo do tempo. Dresner et al. (1996) examinaram o impacto da entrada da Southwest nas companhias aéreas que operam em rotas adjacentes ou em aeroportos próximos àquela entrada pela Southwest; os autores indicaram que, mesmo nesses casos de competição indireta, os preços caíram e o tráfego de passageiros aumentou dramaticamente, ampliando o conhecimento anterior sobre o Efeito Southwest. Um exemplo mais recente é o de Beckenstein e Campbell (2017), que mostraram que a Southwest foi responsável por uma redução média de US$45 no preço das passagens em rotas em que ela entrou no ano de 2016; esse valor representou uma redução de 15% em relação ao preço praticado pelos seus concorrentes, e como consequência dessa redução dos preços médios das passagens aéreas, houve um aumento de 30% no número de viagens aéreas naquelas rotas.

Além de investigar o Efeito Southwest na demanda por transporte aéreo, também existem estudos voltados para as respostas competitivas das companhias aéreas à entrada da Southwest. Ito e Lee (2003), por exemplo, analisaram as respostas das companhias aéreas que atuam utilizando a estratégia de hub-and-spoke à entrada da Southwest nas rotas ligadas aos seus hubs; eles descobriram que, embora algumas companhias aéreas respondam à entrada da Southwest com uma redução acentuada de preços junto com uma agressiva expansão de sua capacidade, a resposta média dos incumbentes foi de se acomodar, alinhando seu preço ao da Southwest e não entrando em guerra de preços com a entrante. Daraban e Fournier (2008) concordaram que os incumbentes reduzem suas tarifas após a entrada da Southwest e também adicionou mais conhecimento à literatura mostrando evidências de que as incumbentes também cortam o preço de suas passagens em antecipação à entrada da LCC, e que as tarifas permanecem mais baixas mesmo após a saída da Southwest do mercado; eles também confirmaram que os preços em uma determinada rota diminuem quando a Southwest serve uma rota adjacente a ela.

Mas a Southwest não é a única LCC do mundo; fundada como Air Southwest em 1967 e adotando o nome Southwest Airlines em 1971, esta empresa se tornou a maior companhia aérea de baixo custo do mundo, servindo como referência para muitos outros LCCs em diferentes mercados. Depois disso, vários estudos foram feitos para analisar e comparar os efeitos dessas novas companhias aéreas de baixo custo no mercado com o Efeito Southwest.

Windle e Dresner (1999) analisaram o impacto da entrada da rota pela Valujet no hub da companhia aérea Delta, e examinaram as mudanças de preço nas rotas que não foram entradas pela LCC; eles descobriram que a Delta reduziu suas tarifas em rotas competitivas conectadas ao seu hub em resposta à concorrência da Valujet, mas não encontrou evidências de um possível aumento de preço em rotas não competitivas para compensar a perda de receita nas rotas competitivas, o que implica que os governos devem incentivar a entrada de LCCs para aumentar o bem-estar do consumidor. Brueckner et al. (2013) analisaram a concorrência das companhias aéreas e os impactos da entrada de low costs nas tarifas aéreas; os resultados mostraram que concorrência com uma LCC tem impactos dramáticos nas tarifas, mesmo que ocorra no par de aeroportos, em aeroportos adjacentes ou mesmo uma concorrência potencial, confirmando e ampliando os resultados anteriores da literatura.

Nos últimos anos, com novas estratégias de mercado e modelos de negócios, algumas companhias aéreas apresentaram um Efeito Southwest maior do que a própria Southwest Airlines. Wittman e Swelbar (2013) descobriram que certas companhias aéreas de baixo custo apresentavam um impacto maior na redução da tarifa média em comparação com a Southwest, e que os especialistas deveriam adotar o nome Efeito JetBlue, que foi uma das LCCs analisadas no estudo.

Nesse contexto dos impactos da presença de uma LCC no mercado da aviação civil, o presente trabalho discutirá o que foi pesquisado recentemente na literatura e como as descobertas mais recentes se contrapõem às conclusões clássicas da literatura sobre o efeito das LCCs em diferentes participantes do mercado de transporte aéreo, que foram divididas em: (i) companhias aéreas concorrentes; (ii) aeroportos; (iii) demanda geral por transporte aéreo; (iv) processo de escolha dos passageiros; e (v) região geográfica.

### III. Impacto das LCCs nas companhias aéreas concorrentes

Existem algumas maneiras de mudar o cenário de qualquer mercado comercial, e uma delas é a entrada de um novo concorrente. Os entrantes podem afetar as empresas existentes tomando seus market shares, reduzindo assim o lucro esperado dessas empresas; numa tentativa de defender a sua fatia de mercado, as incumbentes costumam responder reduzindo seus preços ou mesmo aumentando sua eficiência operacional para reduzir suas perdas (Porter, 2008).

Essa ideia generalista do mundo empresarial se transfere perfeitamente para a indústria da aviação também. Como foi contextualizada anteriormente, a redução do preço das passagens aéreas é uma forma das companhias aéreas responderem à concorrência da Southwest Airlines ou de outras companhias aéreas de baixo custo. Apesar de ser um tema bastante discutido na literatura desde a desregulamentação da aviação civil nos EUA, ainda existem novas pesquisas analisando esse problema por diferentes ângulos.

Alguns dos estudos mais recentes sobre esse tema são de Asahi e Murakami (2017) e Ren (2020), em que ambos revisitaram o impacto de uma LCC no preço das passagens estudando a própria Southwest Airlines. A justificativa para isso é que segundo alguns pesquisadores da área, a Southwest hoje não é mais considerada como uma empresa estritamente low cost (por sua posição dominante no mercado aéreo americano), mas sim uma híbrida de low cost com companhia aérea tradicional, e que, portanto, valeria a pena investigar se ela continua causando os mesmos efeitos nos preços de seus concorrentes como antigamente. Os autores concluem que a Southwest ainda reduz o preço médio das passagens nas rotas em que ela entra, mas essa redução não é tão aguda como fora outrora. Por exemplo, em rotas em que as incumbentes são dominantes ou que possuem um voo direto conectando os principais hubs do país o impacto da Southwest nos preços é mais fraco.

Além de estudos focados na Southwest, também existem estudos que analisaram as LCCs de diferentes mercados, como foram os casos de Chen (2017) e Zhang et al. (2018), que estudaram o caso da China e Austrália, respectivamente. Esses autores concluem que a presença de LCCs nas rotas reduz o preço da passagem, concordando com a literatura clássica, mas que em alguns casos essa redução não é tão acentuada, pois em muitas das rotas analisadas por eles não havia uma presença muito forte de LCCs, diminuindo o seu potencial de redução no preço das passagens.

Bachwich e Wittman (2017) estudaram os efeitos das ultra low cost carriers (ULCCs), que são aquelas companhias aéreas que reduzem ao máximo seu custo operacional, mais ainda do que que low costs clássicas como Southwest, e por esse motivo conseguem oferecer preços ainda menores no mercado. Eles dizem que as tanto as LCCs quanto as ULCCs conseguem reduzir o ticket médio de uma rota, mas que as ULCCs têm um maior poder de redução de preços do que as LCCs. Porém, eles concluem que, devido à estratégia mais agressiva, as ULCCs também tendem a sair mais rápido dos mercados em que ela entrou em comparação com LCCs, não esperando até uma possível maturação de demanda por esse tipo de voo na região.

Ainda sobre preços de passagens, Zou et al. (2017) estudaram o efeito indireto das LCCs sobre o preço do bilhete aéreo ao analisar o preço cobrado por despacho de bagagens. Eles dizem que a presença de uma ULCC ou LCC numa rota ou aeroporto adjacente força os seus concorrentes que cobram por despacho de bagagem a reduzirem o preço das passagens aéreas, corroborando mais uma vez a literatura clássica dos efeitos das LCCs no preço do bilhete.

Além do impacto no preço, como foi levantada na frase de Porter (2008), a presença de um concorrente pode forçar as companhias existentes a mudarem suas operações ou mesmo estratégias de negócio, e essa hipótese também foi testada nos últimos anos pela literatura da aviação. Sun (2015) diz que a presença de LCCs numa rota aumenta a diferenciação dos tempos de partida das aeronaves numa tentativa de evitar a concorrência direta ou o horário de pico, além de oferecer um produto diferenciado aos passageiros. Seguindo nessa linha, Mohammadian et al. (2019) dizem que as companhias aéreas concorrentes chegam a mudar o tipo de aeronave utilizada e até a frequência de voos servidos em uma rota. Já Bendinelli et al. (2016) encontraram resultados sugerindo que a presença de uma LCC pode ser responsável pela internalização dos custos dos atrasos pelas suas companhias aéreas.

Ainda sobre impactos nas operações e estratégias das companhias aéreas incumbentes, Pearson et al. (2015) fizeram uma pesquisa investigativa no mercado asiático para avaliar a capacidade estratégica das companhias aéreas tradicionais para competir com o crescente número de LCCs naquele mercado; eles concluem que o nordeste asiático é a região em que as companhias aéreas estão menos preparada para a concorrência com as low costs, e que portanto as network carriers dessa região devem fortalecer a sua capacidade estratégica tanto para melhorar a sua capacidade de resposta às low carriers, mas também aumentar a sua performance de um modo geral.

E por fim, existem estudos recentes que avaliaram o impacto das LCCs sobre as companhias aéreas que fazem voos fretados. Wu (2016) e Castillo-Manzano et al. (2017) tiveram conclusões semelhantes, e também condizentes com a literatura pré-existente, em que a presença das low costs fez com que o mercado de companhias aéreas charter encolhesse, obrigando essas a mudarem o seu modelo de negócio (ao passar a vender assentos em voos regulares), ou a se tornarem uma low cost também.

## IV. Impacto das LCCs nos aeroportos

Para desenvolver qualquer mercado da aviação, a existência de três peças é extremamente crucial: o aeroporto, a companhia aérea e, é claro, a demanda para usar esse serviço. Pode-se argumentar que aeroportos e companhias aéreas têm uma relação simbiótica, e aqueles que podem usar o potencial máximo de seus parceiros se tornam uma referência a ser seguida por seus concorrentes, seja um aeroporto ou uma companhia aérea. Neste contexto, serão mostradas aqui as pesquisas que analisam o impacto da presença de uma LCC num aeroporto.

Existem trabalhos que tentam relacionar a presença de uma LCC com um possível aumento na eficiência operacional e/ou financeira de um aeroporto. Alguns desses trabalhos, como de Yokomi et al. (2017) e Zuidberg (2017) não encontraram evidências de que a presença de LCCs seja responsável por um aumento na eficiência aeroportuária. Yokomi et al. (2017), por exemplo, concluíram que voos LCCs trazem um menor lucro marginal ao comparar com voos de companhias aéreas tradicionais; enquanto Zuidberg (2017) diz o impacto de uma LCC sobre a lucratividade de um aeroporto é quase não-existente.

Porém, existem trabalhos que encontraram impactos positivos devido à presença de companhias aéreas de baixo custo, como um aumento na eficiência financeira (Button et al., 2017), eficiência operacional (Martini et al., 2020) ou em ambos (Augustyniak et al., 2015). Dessa forma, pode-se perceber que há uma falta de consenso na literatura da área sobre o impacto de uma LCC sobre a eficiência aeroportuária.

Talvez um dos fatores responsáveis por essa discordância na literatura esteja na escolha dos aeroportos analisados. Tavalaei e Santalo (2019) fizeram um estudo analisando as estratégias competitivas dos aeroportos e suas eficiências; eles dizem que aeroportos puros, aqueles que servem exclusivamente voos de companhias aéreas low costs ou tradicionais, são mais eficientes do que aeroportos híbridos, que servem os dois tipos de companhias aéreas.

Além de investigar o efeito das LCCs nas eficiências aeroportuárias existem diversos trabalhos que tentam analisar a relação dessas companhias aéreas com a conectividade de um aeroporto. A conectividade neste contexto é definida como um indicador da concentração de uma rede; portanto, é a capacidade de mover passageiros de um ponto ao outro com o menor número possível de conexões e sem um aumento na tarifa, e que deve se concentrar em atingir tempos mínimos de conexão com o máximo de facilitação, resultando em benefícios para os passageiros.

Zhang et al. (2017) dizem que a presença de uma LCC tem efeito positivo sobre a conectividade aeroportuária, e que isso é especialmente importante para aqueles aeroportos turísticos ou regionais. Essa conectividade pode até ser utilizada como uma forma de capitalização do aeroporto, ao oferecer serviços de facilitação para os passageiros que fazem auto conexão (Chang et al., 2019), que é um segmento do mercado ainda não muito explorado pelos aeroportos, mas que tem um potencial de crescimento no futuro (Cattaneo et al. 2017).

Por fim, Jimenez et al. (2017) estudaram os impactos das LCCs no sistema aeroportuário como um todo, e concluíram que o crescimento no número de companhias aéreas de baixo custo é um importante driver para o desenvolvimento da infraestrutura física dos aeroportos, sendo relacionados à expansão ou construção de novos aeroportos em um mercado.

## V. Impacto das LCCs na demanda por transporte aéreo

Além de impactar nos aeroportos, as companhias aéreas de baixo custo também têm uma forte influência na demanda de viagens aéreas, como foi mostrado em estudos passados do Efeito Southwest. Esses novos estudos da literatura revisitaram o Efeito Southwest ao analisar o potencial gerador de demanda das LCCs no contexto atual, e confirmaram de um modo geral que a presença de LCCs aumenta o número de passageiros num aeroporto (Rolim et al., 2016; Boonekamp et al., 2018), porém em alguns mercados esse potencial ainda é limitado devido à falta de um maior número de voos de companhias aéreas low cost (Valdes, 2015; Tsui e Fung, 2016).

Além de estudos que analisaram o efeito da presença de uma LCC na demanda por voos de um aeroporto/região, também há estudos recentes analisando uma demanda mais específica, a turística. Alsumairi e Tsui (2017) diz que a oferta de voos low cost foram um dos principais fatores para o aumento no número de turistas e mesmo o desenvolvimento do transporte aéreo como um todo na Arábia Saudita. Na mesma linha, Álvarez-Díaz et al. (2019) dizem que a presença de low costs numa região marcada por uma grande diáspora e êxodo de turistas ainda tem capacidade de atração de visitantes, sendo a presença dessas companhias aéreas um fator chave para o desenvolvimento econômico da região.

## VI. Impacto das LCCs no processo de decisão passageiros

Anteriormente foi discutido o papel e os efeitos das companhias de baixo custo na demanda por viagens aéreas. Aqui, a demanda será novamente discutida, mas de maneira granular ao apresentar os estudos sobre como a presença de uma LCC pode afetar a escolha dos passageiros por uma determinada companhia aérea, por determinado aeroporto ou até por um destino de viagem.

Paliska et al. (2016) analisaram como um aeroporto com a presença de voos low cost podem atrair passageiros. Apesar de LCCs normalmente atraírem passageiros que voam a lazer, devido a sua maior sensibilidade ao preço das passagens, nesse estudo foi encontrado que no aeroporto de Trieste, na Itália, a presença de voos low cost atrai mais passageiros business; o principal fator pela preferência desses passageiros pelas LCCs se deve ao fato destas oferecem voos ponto a ponto aos principais hubs da Europa, portanto sem necessidade de conexões. Outro fator curioso do trabalho de Paliska et al. (2016) é que os passageiros que residem em regiões fronteiriças com outros países tendem a utilizar os aeroportos localizados em seu país de origem, e que nesse caso o efeito da presença de LCCs é reduzido.

Além de influenciar na escolha por aeroportos, a presença de uma LCC pode influenciar na escolha da companhia aérea que um passageiro pode tomar segundo estudos recentes. Yang (2016), por exemplo, analisou o caso do nordeste asiático e verificou que os passageiros dessa região têm uma maior preferência por low costs estrangeiras, mostrando que essas LCCs mais conhecidas no mercado têm um impacto maior na demanda se comparadas com as LCCs da região. Já Saffarzadeh et al. (2016) estudaram o mercado iraniano, e concluíram que, embora os passageiros dessa região têm uma preferência por voos low cost, essas companhias aéreas ainda devem oferecer algumas facilidades aos seus passageiros e um bom nível de serviço. Seguindo essa linha do que os passageiros esperam de seus voos, Kim (2015) analisou o efeito do valor percebido entre voos low cost e tradicionais, dizendo que os passageiros são mais criteriosos na avaliação da qualidade do serviço de companhias aéreas low cost, e que mesmo que essa ofereça um bom serviço, não há garantias futuras de que um passageiro volte a voar novamente com a low cost. Por fim, Hunt e Truong (2019) analisaram o caso específico das companhias aéreas de baixo custo de longa distância (do inglês low cost, long-haul – LCLH), mostrando que essas companhias aéreas estão mudando as características do mercado de viagens transatlânticas ao oferecer voos por preços baixos aos seus clientes e ganhando aos poucos espaço no mercado, mesmo com problemas de falta de conforto e conveniência em seus voos.

Existem também estudos recentes que analisaram o efeito da presença de uma companhia aérea low cost em diferentes perfis de passageiros. Cattaneo et al. (2016), por exemplo, analisaram o caso dos estudantes na Itália, especificamente se a presença de uma companhia aérea low cost em um aeroporto pode ser um fator decisivo na escolha da universidade que eles estudarão no futuro; os resultados mostram que a resposta para essa pergunta é sim, que é um fator importante para os estudantes a presença de um aeroporto que serve voos low costs perto da universidade. Já Clavé et al. (2015) estudaram o processo de tomada de decisão dos turistas e como eles escolhem os destinos que irão viajar; segundo eles, a presença de uma low cost pode sim ter uma influência na escolha do destino, mas que nem todos os turistas desse destino o escolheram devido à presença de LCCs. Borhan et al. (2017) estudaram como a presença de companhias aéreas low cost podem influenciar os motoristas automotivos da Líbia a trocarem de modal de transporte, mostrando que a presença de voos low costs podem ser um fator importante para que esses passem a utilizar mais o modal aéreo.

E por fim, Valdes e Gillen (2018) estudaram como as LCCs podem impactar no bem-estar social dos passageiros. Segundo a teoria clássica, a mera presença de uma LCC deveria aumentar o bem-estar social devido ao efeito redutor de preço das passagens aéreas. Eles afirmam que em diversos cenários previstos, a presença de voos LCCs num aeroporto elevou, de fato, o bem-estar da população, mas também existiram cenários em que isso não ocorreu devido ao abuso de mercado que as companhias podem exercer num posição dominante, e que, portanto, para garantir que uma LCC eleve o bem-estar social, é necessário fazer uma análise detalhada do mercado e não simplesmente incentivar essas companhias aéreas a entrarem em rotas.

## VII. Impacto das LCCs em uma região geográfica

Além de analisar o efeito das LCCs nas decisões dos passageiros, existem estudos sobre como uma LCC pode ter impacto em determinadas regiões sem o foco no turismo. Bowen Jr (2016), por exemplo, estudou como a presença de uma companhia aérea de baixo custo no sudeste asiático pode aumentar a acessibilidade de uma região periférica ao mercado global do transporte aéreo, concluindo que as LCCs têm um papel principal para o crescimento desses aeroportos secundários e a toda região servida por eles. Taumoepeau et al. (2017) seguiram na mesma linha e estudaram o mercado aéreo da Oceania; por essa ser uma região com características distintas (economia pequena, menor número de habitantes e comunidades mais isoladas), a presença de uma LCC híbrida pode ser a chave para o desenvolvimento da aviação nessa região.

## VIII. Considerações Finais

E afinal, por que devo me importar com as companhias aéreas de baixo custo? Como pôde ser visto neste trabalho, em um setor dinâmico como o transporte aéreo, uma nova entrada de companhia aérea pode abalar o cenário existente nesse mercado; porém, quando essa nova companhia aérea possui um modelo de negócios de baixo custo, alguns efeitos dessa entrada podem ser amplificados e sentidos por diferentes participantes no mercado de transporte aéreo. As pesquisas clássicas sobre os impactos de uma entrada na LCC avaliaram principalmente como essas empresas podem afetar as tarifas aéreas e a demanda de voos em uma região. Com o sucesso da *Southwest* Airlines no mercado americano, esse modelo de negócios foi replicado ao redor do mundo, passando até por transformações, ao mesmo tempo em que novas companhias aéreas, com uma nova visão do modelo de baixo custo, entraram no mercado. Nesse contexto de uma indústria extremamente dinâmica, o presente trabalho compilou e discutiu o que há de novo na literatura da área sobre os efeitos da LCC nos diferentes *stakeholders* do mercado do transporte aéreo, sendo eles: concorrentes, aeroportos, demanda geral, passageiros e regiões geográficas.

Com relação aos efeitos da LCC nas companhias aéreas rivais, estudos demonstraram que uma companhia aérea de baixo custo força uma resposta de seus rivais para não perder sua posição dominante no mercado, seja reduzindo o preço das passagens aéreas ou mesmo adaptando suas operações para obter uma melhor eficiência. No que diz respeito às companhias aéreas de voos fretados, estudos recentes confirmaram os resultados existentes na literatura indicando um efeito de substituição das companhias aéreas fretadas por LCCs.

O presente trabalho também mostrou que os efeitos nos aeroportos ainda não são claros e que uma possível direção para pesquisas seja investigar os efeitos da LCCs na receita e conectividade do aeroporto. Além disso, foi visto que as LCCs ainda aumentam a demanda geral de transporte aéreo, mostrando que o Efeito *Southwest* ainda é válido atualmente, seja para o público em geral ou para rotas turísticas.

Além disso, vários estudos analisados aqui mostraram que a presença LCCs é um fator importante para a escolha dos passageiros, seja dentro do mercado de viagens aéreas, como companhias aéreas e aeroportos, ou externo a esse mercado, como universidades e destinos turísticos; e que as LCCs podem induzir o desenvolvimento de regiões não atendidas pelo serviço de transporte aéreo.